\documentclass[aps,
groupedaddress,showpacs,nofootinbib,prl,twocolumn]{revtex4}
\usepackage{amssymb,amsmath}
\usepackage{graphicx}
\usepackage[english]{babel}
\usepackage{graphicx}
\usepackage{dcolumn}

\usepackage{bm}
\def\nablab{\mbox{\boldmath$\nabla$}}
\def\nablabs{\mbox{\scriptsize\boldmath$\nabla$}}
\def\nablabf{\mbox{\footnotesize\boldmath$\nabla$}}
\def\etab{\mbox{\boldmath$\eta$}}
\def\etabs{\mbox{\scriptsize\boldmath$\eta$}}

\def\dx{d^D\!x}
\def\fsz{\tiny}

\def\aut#1{#1}

\def\ins#1{{\it #1}}
\def\sdag{\dagger}

\def\meq#1{}
\def\meq#1{}

\def\ins#1{}

\def\comment#1{}

\def\dst{\displaystyle}
\def\mn#1{\marginpar[]{\scriptsize#1}}
\def\mn#1{}

\def\IncludeEpsImg#1#2#3#4{\renewcommand{\epsfsize}[2]{#3##1}{\epsfbox{#4}}}

\usepackage{ulem}
\usepackage{cancel}
\usepackage{color}
\def\rr#1{\textcolor{red}{#1}}

\def\mn#1{\marginpar[\tiny{\rr{#1}}]{\tiny{\rr{#1}}}}

\def\rr#1{}

\usepackage{hyperref}

\def\IncludeEpsImg#1#2#3#4{\renewcommand{\epsfsize}[2]{#3##1}{\epsfbox{#4}}}

\newcommand{\be}{\begin{equation}}\newcommand{\ee}{\end{equation}}
\newcommand{\bea}{\begin{eqnarray}}\newcommand{\eea}{\end{eqnarray}}
\newcommand{\beaa}{\begin{eqnarray}}\newcommand{\eeaa}{\end{eqnarray}}
\newcommand{\ba}{\begin{array}}\newcommand{\ea}{\end{array}}
\newcommand{\bit}{\begin{itemize}}\newcommand{\eit}{\end{itemize}}
\newcommand{\ben}{\begin{enumerate}}\newcommand{\een}{\end{enumerate}}

 \newcommand{\sfrac}[2]{\raisebox{0.095ex}{\scriptsize${\frac{#1}{#2}}$}}

									\def\be{\begin{equation}}
\def\ee{\end{equation}}				\def\bea{\begin{eqnarray}}				\def\eea{\end{eqnarray}}
\def\bear{\begin{array}}				\def\eear{\end{array}}

												\def\x5{x^{5}}

\begin{document}

\title{Fractional %
Quantum Field Theory,
Path Integral,\\ and Stochastic Differential Equation
for Strongly Interacting Many-Particle Systems}

\comment{
\author{H. Ruby}
\email{hruby@live.de}
\affiliation{ICRANeT Piazzale della Repubblica, 10 -65122, Pescara, Italy}
}

\author{Hagen Kleinert}
\email{h.k@fu-berlin.de}


\affiliation{Institut f{\"u}r Theoretische Physik, Freie Universit\"at Berlin, 14195 Berlin, Germany}
\affiliation{ICRANeT Piazzale della Repubblica, 10 -65122, Pescara, Italy}


\vspace{2mm}
\def\sch{Schr\"odinger}
\def\comment#1{}

\begin{abstract}

While free and {\it weakly\/} interacting particles
are well described by a second-quantized
nonlinear 
Schr\"odinger field, or relativistic versions
of it,  with various approximations, the fields of {\it strongly\/} interacting
particles are governed
by {\it effective
actions\/}, whose quadratic terms are extremized
by 
fractional wave equations.
Their particle orbits perform  universal L\'evy walks
rather than Gaussian random walks with perturbations.

\end{abstract}

\pacs{95.35+d.04.60,04.20C,04.90}

\maketitle

Quantum-mechanical physics is
explained
with high accuracy
 by
\sch{} theory. The wave equation
for  many particles can
convenienty be reformulated as
a second-quantized {\it field theory\/},
with an action
that is the sum of quadratic 
and an interacting term
\begin{eqnarray}
 {\cal A}= {\cal A}_{2}+ {\cal A}_{\rm int},
\label{4.28A}\end{eqnarray}
where the term ${\cal A}_{2}$ has typically 
the form
\begin{eqnarray}\!\!\!\!\!\!\!\!\!\!\!\!\!\!\!\!
{\cal A}_{2}\!=\!\! \int \!\!\dx dt  \psi  ^{*}  ({\bf x},t)[i\partial_t
\!+\!\hbar ^2  \nabla^ 2/2-\!V({\bf x})
]\psi  ({\bf x},t),\!\!\!\!\!\!\!\!\!\!\!\!\!\!\!
\label{4.28B}\end{eqnarray}
with $D$ being the space timension, 
$m$ the mass, and $V({\bf x})$ some external potential.
The interaction term
$ {\cal A}_{\rm int}$ may be approximated 
in molecular systems by a
fourth-order term 
in the field
\begin{eqnarray}
\frac{1}{2}\!
     \int \hspace{-1pt}\hspace{-1pt}\!\dx \dx'dt \psi ^{*}  ({\bf x}',t)
     \psi  ^{*}  ({\bf x},t)V_{12}({\bf x},{\bf x}')
\psi  ({\bf x},t)
     \psi  ({\bf x}',t),
\label{@3}\end{eqnarray}
where $V_{12}({\bf x},{\bf x}')$ is some two-body potential.

If relativistic velocities
are present, the
field
is generalized to
a scalar
Klein-Gordon field, or a
quantized
Dirac field. In molecular physics, the fourth-order term is due to the exchange 
of
a
 minimally coupled  quantized photon field and is proportional to 
$e^2$,
where $e$ is the electric charge.
The field equations may be studied 
with any standard method of quantum field theory, and 
corrections can be derived using perturbation theory
in powers of $\alpha\equiv e^2/\hbar \approx 1/137$.
Since $\alpha$ is very small,
this appeoch is quite successful.

If time is
continued analytically to imaginary values $t=i\tau$,
one is faced with the so-called Euclidean
version of quantum field theory.
Then perturbation theory
may be understood as developing
a theory of particle physics
from
an expansion around Gaussian
random walks.
Indeed, the relativistic scalar free-particle propagator of mass $m$
in $D+1$-dimensional  euclidean
energy-momentum space $p^\mu=({\bf p},p_4)$,
has the form
\begin{eqnarray}
G({p})=\frac1{{\bf p}^2+ p_4^2+m^ 2}
=\int_0^\infty ds\, e^{-s m^2}
 e^{-s ({\bf p}^2+ p_4^2) },
   \label{@P}
\end{eqnarray}
where
the energy has been continued analytically to $p_4=-iE$.
The Fourier
transform of $ e^{-s ({\bf p}^2+ p_4^2) }
$
is the distribution of  Gaussian random walks
of length $s$
in $D+1$ euclidean
dimensions
\begin{eqnarray}
P({\bf x},x_4)=
(4\pi s)^{-(D+1)/2}
e^{-(|{\bf x}|^{2}+x_4^2)/4s }
,
\label{@}\end{eqnarray}
which makes
the propagator (\ref{@P})
a superposition of such walks
with lenghts distributed
like
$e^{-s m^2}$ \cite{FO,Feynman,PI}.
This propagator
is the relativistic version of the
free-field propagator of
the action (\ref{4.28B}).
The second-quantized field theory
described
by
(\ref{4.28A})
accounts for
grand-canical ensembles
of orbits with their two-body  interactions \cite{DO}.

Gaussian random walks
are a natural and rather universal
starting point for
many stochastic processes.
For instance, they form
the basis of the most important tool
in the theory of financial markets,
the Black-Scholes option price theory \cite{BLS} (Nobel Prize 1997),
by which a portfolio of assets
is hoped to remain steadily growing through hedging.
In fact, the famous {\it central-limit theorem\/}
permits us to prove that
many independent random movements
of finite variance
always pile up to display
a Gaussian distribution
 \cite{BP}.

However, since the last stock market crash and the
still ongoing
financial crisis
it has become clear
that realistic
distributions
belong to a more general universality class,
the so-called  L\'evy stable distribution.
They are the univarsal results
of a
pile up of
random movements
of infinite variance \cite{remn}. They
account
for the fact that
rare events, which
initiate crashes,
are much more frequent
than in Gaussian distributions.
These are events in
the so-called  L\'evy tails
$\propto 1/|x|^{1+\lambda}$ of the distributions, 
whose
description
requires a Hamiltonian  \cite{PI}
\begin{eqnarray}
H={\rm const}\,(p^2)^{\lambda/2}.
\label{@}\end{eqnarray}
Such tail-events are present
in the
self-similar
 distribution of matter in the universe \cite{EI,VS,DU}, 
in velocity distributions of many body sytems with long-range forces 
\cite{RUFFO}, and
in the distributions
of windgusts 
\cite{PEINKE}, oceanic moster waves \cite{HOFF}, 
and
earthquakes
\cite{SEISM},
with often catastrophic consequences.
They are a consequence
of rather general maximal entropy assumptions \cite{GMTS}. 
In the limit $\lambda\rightarrow 2$, the  L\'evy  distributions
reduce to Gaussian  distributions.

\comment{
Another example is the distribution of wind gusts
which may destroy a good part of the windmills
that are supposed to replace
the nuclear reactors in the supply of
energy
in our civilization \cite{PEINKE}.
A third example is the statistics of earthquakes
\cite{SEISMI}.
}

The purpose of this note is to point out, that
such distributions
occur quite naturally also in
many-particle systems, provided the interactions are very strong \cite{STR}.
They have been observed
in numerous experiments
at second-order phase transitions.
The most accurate measurement of this type
was done in a satellite (the so called Infrared Atronomical Satellite IRAS)
by
studying
the singularity of the specific heat
of
superfluid
${}^4$He near the critical temperature  \cite{rLipa}.
The observation agreed extremely well
with the theoretical strong-coupling prediction
\cite{KHE}.

The field of a strongly interacting $N$-body system
is
usually
a multivalued function. Singularities
perforate the space via vortex lines
(for instance in type II
superconductors or in superfluid ${}^4$He),
or via
line-like defects in the
displacement field of
a world-crystal
formulation of
Einstein(-Cartan) gravity \cite{MVF}.
If the positions of two
particles are exchanged, one obtains
a factor $+1$ for bosons or $-1
$ for electrons.
In two dimensions, one may even obtain
a general phase $e^{i\phi}$ (anyons) \cite{BPZ}.

A strongly interacting field system
has
a conformally invariant
Green function
  \cite{ABAR,KS22,BPZ}
\begin{eqnarray}
G({\bf p},p_4)=[ p_4^{1-\gamma}\phi({\bf p}^2/ p_4^{z})]^{-1}.
\label{@Prop}\label{@6}
\end{eqnarray}
If the dimension $D$ differs
 only by
 a very small amount $\epsilon$
 from   the critical dimension $D_c$, where the theory is scale-invariant,
i.e., $D=D_c+\epsilon$,
 then
 $\gamma$ is of order $\epsilon$
and
$z$  differs from unity by a similar amount.
Such a power behavior is assured near $D_c$
if the Gell-Mann-Low function \cite{GML} has an infrared-stable 
fixed point in the renormalization 
flow of the coupling constant.
Very close to the critical dimension, a lowest approximation to
$
G({\bf p},p_4)$
is
\begin{eqnarray}
G({\bf p},p_4)=\{p_4^{1-\gamma}[1+D_\lambda(
{\bf p}^2/p_4^{z}
)^{\lambda/2}]\}^{-1},
\label{@PropaGAL}\end{eqnarray}
where $\lambda$ is close  to 2, and $D_\lambda$ is a
generalization of the diffusion constant
in the Fokker-Planck equation.

Time-independent propagators
involve
the limit
$p_4\rightarrow 0$,
where the correlation function behaves like
\begin{eqnarray}
G({\bf p},0)\propto
|{\bf p}|^{-2+\eta}.
\label{@PropaGAL2}\end{eqnarray}
The index $\eta$ is the {\it anomalous dimension\/} of the field,
which is also of order $\epsilon$.
The existence of this limit in (\ref{@PropaGAL})
fixes the scaling relation
\begin{eqnarray}
\lambda=2(1-\gamma)/z=2-\eta.
\label{@}\end{eqnarray}
See Appendix for the calculation of the exponents to order $\epsilon$.
The Green function (\ref{@PropaGAL})
determines
the probability
distribution
of particle after a time $t$
via the {\it double
fractional Fokker-Planck equation\/}
\begin{eqnarray}
[\hat{p}_4^{1-\gamma}+D_\lambda(\hat{\bf p}^2)^{\lambda/2} 
]P({\bf x},t)=\delta(t)\delta^{(D)}({\bf x})
,
\label{@10}\end{eqnarray}
where
 $\hat{p}_4\equiv \partial_t$,
$\hat {\bf p}\equiv i\partial_{\bf x}\equiv i \nablab$.
A convenient definition of the fractional derivatives
uses the same formula
as in the dimensional 
continuation of Feynman diagrams
$(-\nablab^ 2)^{\lambda/2}=\Gamma[\lambda/2]^{-1}\int d\sigma \sigma^{-\lambda/2-1}
e^{\lambda \nablabs^ 2/2}$ \cite{METZ,METZz}.
\comment{
$\hat p_4^{1-\gamma}$ and $(\hat {\bf p}^2)^{\alpha/2}$
$\hat p_4^{1-\gamma}f(t)
\equiv \Gamma^{-1}[1-\gamma]\int_t^\infty dt'(t-t'+i\epsilon)^{-2+\gamma}
f(t')$ \cite{RAI}.
The $D$-dimensonal operaton
$\hat p_4^{\lambda/2}f({\bf x})$}%
The solution of (\ref{@10})
is given in the literature
 \cite{DUAN} and
reads
\def\tau{t}
\begin{eqnarray}
\!\!\!\!\!\frac{\tau^{-\gamma}}{\pi^ {D/2}|{\bf x}|^{D/2}}H^{2,1}_{2,3}
\left(\!\frac{|{\bf x}|^{\alpha}}{2^\alpha D_\lambda\tau^{1-\gamma}}\bigg |
^{(1,1);(1-\gamma,1-\gamma)}
_{(1,1),(D/2,\alpha/2);(1,\alpha/2)}\!
\right)\!,
\label{@11}\end{eqnarray}
where $H^{2,1}_{2,3}$ is a {\it Fox H-function\/} \cite{FH}.%
\comment{
The general properties of fractional differential operators
are discussed in \cite{LEVY}.
}
In the limits
 $\gamma=0$ and $\alpha =2$, this reduces to  the
standard quantum mechanical Gaussian expression
$(4\pi D_\lambda \tau)^{-D/2}
e^{-|{\bf x}|^{2}/4D_\lambda\tau }
$.
For $\gamma=0,\alpha =1$,
the result is
\begin{eqnarray}\!\!\!\!
P({\bf x},t)=\frac{D_\lambda\tau}{\pi ^{(D+1)/2}|{\bf x}|^{D+1}}
H^{1,1}_{1,1}
\left(
\!{\frac{D_\lambda^2\tau^{2}}{|{\bf x}|^2 }} \bigg |
^{(1/2-D/2,1)}
_{(0,1)}\!
\right)\!,
\label{@13A}\end{eqnarray}
which is simply
the Cauchy-Lorentz distribution function
$$[\Gamma(D/2+1/2)/\pi^{(D+1)/2}]D_\lambda\tau/[(D_\lambda\tau)^2+|{\bf x}|^2]^{D/2+1/2}.$$%
\comment{
Their
treatment in terms of Gaussian random walks
has been developed in \cite{Feynman}, and its generalization
to  L\'evy walks
was discussed in \cite{PI}.
}

The probability (\ref{@10})
may
 be calculated from the
 {\it doubly fractional canonical path integral\/}
over
 fluctuating orbits $t(s), {\bf x}(s)$
 $p_4(s),{\bf p}(s)$
viewed as functions of some
pseudotime $s$ \cite{REM1}:
\begin{eqnarray}
\{  {\bf x}_b  t_b s_b| {\bf x}_a t_s s_a\}=\int
{\cal D} {\bf x}{\cal D}\tau
 {\cal D} {\bf p}{\cal D}p_4
e^ {-{\cal A}},
\label{@CFP}\end{eqnarray}
where ${\cal A}$ is the euclidean action
of the paths $t(s),{\bf  x}(s)
$:
\begin{eqnarray}
{\cal A}=\int ds[i( {\bf p} {\bf x}'-i
p_4t')-{\cal H}({\bf p},p_4)].
\label{@}\end{eqnarray}
Here
 $t'(s)\equiv dt(s)/ds$, ${\bf x}'(s)\equiv d{\bf x}(s)/ds$,
and
${\cal H}({\bf p},p_4)=p_4^{1-\gamma}+D_\lambda(\hat{\bf p}^2)^{\lambda/2}$.
At each $s$,
the integrals over the components of
 ${\bf p}(s)
$ and $p_4(s)$
run
from $-\infty$ to $\infty$,
whereas
those over
 $p_4(s)$
run
from $-i\infty$ to $i\infty$.
At the end we obtain $P({\bf x},t)$ from
the integral
$\int_0 ^{\infty}ds \{ {\bf x} \, t\,s| {\bf 0}\,0 \,0\}$.

If $\gamma=0$,
the path integral over
$p_4(s) $ yields the functional
$\delta[\tau'(s)-1]$, which
 brings
(\ref{@CFP})
to the canonical
path integral
\begin{eqnarray}
({\bf x}_b t_b|{\bf x}_a t_a)=\int
 {\cal D} {\bf x}
{\cal D} {\bf p}
e^ {-{\cal A}'},
\label{@}\end{eqnarray}
with
\begin{eqnarray}
{\cal A}'=\int d\tau [i{\bf p} \dot {\bf x}-D_\lambda(\hat{\bf p}^2)^{\lambda/2}
].
\label{@}\end{eqnarray}
Now
$P({\bf x},t)=({\bf x} t|{\bf 0}\,0)$
satisfies
the ordinary
fractional
Fokker-Planck
equation
\begin{eqnarray}
[\hat p_4+
D_\lambda(\hat{\bf p}^2)^{\lambda/2}
]P(t,{\bf x})
=\delta(t)\delta^{(D)}({\bf x})\label{eq14X}.
\label{@FSE}\end{eqnarray}
This has been discussed
at length
in  recent literature \cite{LASKIN}.

At this place it is worth
mentioning that
the probability
can be written as
a superposition $
\int_0^\infty( d\sigma/\sigma)
 f_\lambda(\sigma t ^{-2/\lambda})P_{\rm G}(\sigma,{\bf x})$
 of Gaussian
distributions
$ 
P_{\rm G}(\sigma,{\bf x})=
{(4\pi \sigma)^{-D/2}}e^{-{\bf x}^2 /4\sigma}$
with weight
\begin{eqnarray}
f_\lambda(\sigma)=
S_D\sum_{n=1}^\infty
\frac{(-1) ^ n{\sigma^{-n \lambda/2}}}{(n+1)!\Gamma(D-1-n \lambda/2)}D_\lambda^{n/\lambda},
\label{@}\end{eqnarray}
where $S_D=2\pi^{D/2}/\Gamma(D/2)$ is the surface of a sphere in $D$ dimensions.

If $\gamma\neq0$,
the above functional $\delta$-function is softened, and
the relation between the pseudotime $s$ and the physical time
becomes
stochastic. It is governed
by the probability  distribution
that solves the path integral
the
\begin{eqnarray}
\{t_b s_b|t_a s_a\}=\int {\cal D} {t} {\cal D} {p_4}
\exp\left \{\int ds\,[p_4 {t}'-p_4^{1-\gamma})]\right\}.
\label{@}\end{eqnarray}
For imaginary $p_4=-iE$,
we define a {\it noise Hamiltonian\/} $\tilde H(\eta)$
 which has the property
that \cite{REM1,REMAN}
\begin{eqnarray}
e^{-p_4^{1-\gamma}}=\int_{-\infty}^{\infty} d \eta e^{-p_4\eta -\tilde H(\eta)}.
\label{@18}\end{eqnarray}
The inverse of the
Fourier   integral yields the
{\it noise probability\/}
$P(\eta)=
\int_{-i\infty}^{i\infty}d p_4e^{
p_4
\eta
-p_4^{1-\gamma}}$, and a probability functional
 \cite{REM2}:
\begin{eqnarray}\!\!\!\!P[\eta]
\equiv  e^{-\int ds \tilde H(\eta)}\!=\!
\int{\cal D} p_4\exp\left[\int ds\,(
p_4
\eta
-p_4^{1-\gamma})
\right]\!.
\label{@21}\end{eqnarray}
Using this we may solve
the
stochastic differential equation of the
Langevin type
\begin{eqnarray}
t'(s)=\eta(s),
\label{@19}\end{eqnarray}
in which the noise $\eta(s)$
has a zero expectation value for each $s$,
and the correlation functions for $n=2,4,6,\dots~$:
\begin{eqnarray}
\comment{
\langle \eta(s_1)
\eta(s_{2})\rangle& \equiv&
\int {\cal D}\eta\,\eta(s_1)
\eta(s_{2})P[\eta]
, \\
&\vdots&\nonumber \\
}
\langle \eta(s_1)\dots
\eta(s_{2n})\rangle &\equiv&
\int {\cal D}\eta\,\eta(s_1)\dots
\eta(s_{2n})P[\eta].
\label{@}\end{eqnarray}

If $\gamma=0$, the solution of (\ref{@21})
is $P[\eta]=\delta[\eta(s)-1]$,
implying that $\eta(s)$ ceases
to fluctuate, and (\ref{@19})
becomes
$t'(s)\equiv1$, so that
$t\equiv s$.

In the past, many nontrivial
\sch{} equations (for instance that of the $1/r$-potential)
have been solved with  path integral methods
by re-formulating them on the
pseudotime axis $s$,
that is related to the time $t$ via a {\it space-dependent
differential equation\/}
$t'(s)=f(x(t))$. This method,
invented by Duru and Kleinert
\cite{DK} to solve the path integral of the hydrogen atom,
has recently been applied successfully
to various
Fokker-Planck equations \cite{DW,TEMP}.
The stochastic differential equation
(\ref{@19})
may be seen as a stochastic version of the Duru-Kleinert transformation
that promises to be a useful tool to study
non-Markovian systems.

Certainly,
the solutions of Eq.~(\ref{@FSE})
can also be obtained from  a stochastic differential equation
\begin{eqnarray}
\dot{\bf x}=\etab,
\label{@}\end{eqnarray}
whose noise is distributed with a 
fractional probability
\begin{eqnarray}
P[\etab]=\int\! {\cal D}^D\!x e^{
\int\! dt(i{\bf p}\cdot 
\etabs
-D_\lambda({\bf p}^2)^{\lambda/2})
}.
\label{@}\end{eqnarray}

\comment{
Note that there is a stochastic relation between the time $\tau$ and the
pseudotime $s$ for which the \sch{}r equation
is valid. It is found
by solving
the stochastic differential equation
\begin{eqnarray}
\tau'=\eta(s),
\label{@}\end{eqnarray}
where the noise has the distribution
\begin{eqnarray}P[\eta]
\equiv\int{\cal D} E\exp\left\{\int ds\,[
iE
\eta
-E^{1-\gamma}]
\right\}\!.
\label{@}\end{eqnarray}
On the $s$-axis, the amplitudes
are Markovian.
On the $\tau$-axis, this is no longer true.
}

Experimentally, a system with
in the strong-coupling limit can be produced
by forming a Bose-Einstein condensate  (BEC)
in a magnetic field whose strength is
tuned to a Feshbach resonance \cite{ZW} of the two-particle
interaction.
In a BEC, the four-field term in the
interaction (\ref{@3})  is local
and parametrized by
$V_{12}({\bf x},{\bf x})\propto g\delta({\bf x}-{\bf x}')$.
At the Feshbach resonance, the bare coupling strength
$g$ goes to infinity \cite{REMST}, and the renormalized coupling
$g_R$, multiplied by $6\mu^{-\epsilon}
/ (4\pi)^2$,  
 converges to a fixed point $g^*\approx 0.503$\, [see 
Fig. 17.1 in Ref.~\cite{KS22}),
where $\mu$ is some mass scale.

The theoretical tool to describe the physics  in this regime is
the effective action $\Gamma[\Psi,\Psi^*]$.
  This a functional of the classical expectation values of the quantum fields
$\Psi(t,{\bf x})\equiv \langle \psi(t,{\bf x})\rangle$,
and contains all information of
the full quantum 
theory  \cite{KS22,HEB}.
It
is the Legendre transform of
the generating functional $Z[\eta,\eta^*]=\int {\cal D}\psi {\cal D}\psi^*
e^{-{\cal A}-\eta^*\psi-\eta\psi^*}$ of the full quantum theory, and is extremal
on the physical field expections. All its vertex
functions can be found from the functional derivatives of $\Gamma[\Psi,\Psi^*]$.
In the strong-coupling limit,
the effective interaction changes the interaction (\ref{@3})
to 
an anomalous power law
$
\Gamma^{\rm int}[\Psi,\Psi^*]
=(g_c/2)\int d\tau \dx\, |\Psi(\tau,{\bf x})|^ {\delta+1}$. The power $\delta$
is a
critical exponent
that is measured experimentally by the relation $B=|\Psi|^\delta$.
Its value is determined 
by $\eta$ via the so-called hyperscaling relation \cite{REMHS}
$\delta=(D+2-\eta)/(D-2+\eta)
.$
The  value of $g_c$ is related to 
the critical value $g^*\approx 0.503$ by
$g_c \mu^{-\eta D/(D-2+\eta)}
=(2 g^*)^{(\delta-1)/2} (4\pi)^2/24\approx 6.7.$
As a possible application we may study 
the behavior of
a triangular lattice
of vortices
which form in a rotating Bose-Einstein condensate
\cite{DALI}, and letting the
magnetic field approach a Feshbach resonance.

The results
may then be
compared with  a calculation based on a new
field
equation
that generalizes the 
famous 
Gross-Pitaevskii equation 
\cite{GRP}
\begin{eqnarray}
\left[\hat E\!-\frac{1}{2m}
\hat{\bf p} ^2\! -
g
|\Psi(\tau,{\bf x})|^{2
}\right]\Psi(\tau,{\bf x})=0.
\label{@GP}\end{eqnarray}
The new equation is obtained by
extremizing the effective action
$\Gamma[\Psi,\Psi^*]=
\Gamma_0[\Psi,\Psi^*]+
\Gamma^{\rm int}[\Psi,\Psi^*]
$,
where
\begin{eqnarray}\!\!\!
\Gamma_0
\equiv
\int d\tau \dx\,\Psi^\sdag(\tau,{\bf x})
[\hat E^{1-\gamma}-D_\lambda
(\hat{\bf p} ^2)^{\lambda/2}]\Psi(\tau,{\bf x})
.
\label{@23}\end{eqnarray}
\comment{
and
\begin{eqnarray}
\Gamma^{\rm int}[\Psi,\Psi^*]
=g_c \int d\tau \dx\,
|\Psi(\tau,{\bf x})|^{\delta+1}.
\label{@}\end{eqnarray}
}By forming $\delta{\cal A}^{\rm eff}/\delta \Psi^\sdag(\tau,{\bf x})
$,
we obtain
what may be called
the
{\it fractional
Gross-Pitaevskii equation\/}:
\begin{eqnarray}\!\!\!\!\!
 \left[\!\hat E^{1-\hspace{-1pt}\gamma}\!\!-\!D_\lambda\hspace{-0pt}
(\hat{\bf p}^2\hspace{-1pt}) ^{1\hspace{-1pt}-\eta/2}\!\! -\!\frac{\delta\!+\!1}{4
\mu^{\eta}
}
g_c
|
\Psi(\tau,{\bf x})|^{\delta-1}\!\right]\!\!\Psi(\tau,{\bf x})
\!=\!0.
\label{@fGP}\end{eqnarray}

The fractional \sch{}
equation has many problems, such as
the nonvalidity of the quantum superposition law,
the violation of unitarity of the time evolution,
and the violation of probability conservation
which can produce nonsensical probabilities $>1$ \cite{LASKIN}.
However, these problems exist only
if we restrict ourselves
to
the free  
effective
action  (\ref{@23}), but this
is meaningless, since
the
entire theory
is only defined by the effective action
in the strong-coupling limit --- and this
contains  necessarily additional nonquadratic terms.
Hence it does not possess
free quasiparticles as in
the time-honored Landau theory of Fermi liquids \cite{SCH}.
There is always an interaction
that invalidates the standard discussion
of \sch{} equations.
In fact, the theory of high-$T_c$ superconductivity
must probably be built
as a true strong-coupling theory of this type with
electrons being non-Fermi liquids  \cite{SCH}.

The relativistic version
of the entire discussion is
simpler
since it is based on the euclidean Green function
(\ref{@PropaGAL2}) in which ${\bf p}$
denotes the $D-1$-dimensional vectors $({\bf p}, p_4)$.
The
 Fourier transform
is the distribution
fulfilling the Fokker-Planck equation
\begin{eqnarray}
[\partial_s+(\hat {\bf p}^2)^{1-\eta/2}]P(s,\hat {\bf x})=\delta(s)\delta^{(D+1)}({\bf x}).
\label{@}\end{eqnarray}
and possessing
the path integral representation
\begin{eqnarray}
P(s,\hat {\bf x})=\int
 {\cal D} {\bf x}
{\cal D} {\bf p}
e^ {\int ds [i{\bf p} \dot {\bf x}-(\hat{\bf p}^2)^{1-\eta/2}]}.
\label{@}\end{eqnarray}
The $\epsilon$-expansion is now around $D_c=4$
in powers of $\epsilon=-(D-D_c)$. The critical exponent
 $\eta$ is small of order $\epsilon^2$: $\eta=\epsilon^2/50+\dots\approx 0.04 $.
It can be ignored for $\epsilon=1$.
The power $  \delta$ in the interaction
is $3+\epsilon+23 \epsilon^2/50+\dots\approx 4.76$ \cite{REMAR}.

\hyphenation{
e-qua-tion}
The time-independent
{\it fractional
Gross-Pitaevskii equation\/} reads now
\begin{eqnarray}\!\!\!\!\!\!
\left[
(\hat{\bf p}^2) ^{1-\eta/2}\! +\frac{\delta\!+\!1}{4\mu^{\eta}
}
g_c
|\Psi({\bf x})|^{\delta-1}\!\right]\!\Psi({\bf x})=0, 
\label{@GPP}\end{eqnarray}
 with
$g_c\approx 27$.
For a $d=D-1$ -dimensional vortex in $D=3$ dimensions, it is 
solved by $\tilde\Psi({\bf x})=a|{\bf x}_\bot |^{-A}$
with $A=(2-\eta)/(\delta-1)=D/2-1+\eta/2\approx1/2 $ 
and for $\mu=1$: $[(\delta+1)a^{\delta-1}/4]g_c=-
{}^{d}\!\hspace{1pt}c_{\lambda+A-d}{}^{d}\!\hspace{1pt}c^{-1}_{A-d}\approx 0.2$, 
$\lambda=2-\eta$
 \cite{METZ}.

\comment{
In two dimensions,
a list of possible
critical exponents
has been derived in Ref.~\cite{REFBPZ}
from group theory.
}

To compare our theory with experimental data,
we must study the BEC  
in  the scale-invariant
strong-coupling limit. 
This is reached either by going to the temperature $T_c$ of
the second-order phase transition,
or by raising the magnetic field $B$ towards
the field strength $B_c$ of a Feshbach resonance.
Then the coherence length $\xi $ grows
like $\xi\propto |t|^{-\nu}$ where
$\nu\approx 2/3$ \cite{KS22,DR}, and
 $t\equiv 1-T/T_c$
or  $t\equiv 1-B/B_c$  (\cite{ZW}).
If the BEC is enclosed in a weak harmonic trap,
this adds in the brackets
of (\ref{@GP}) a term $\propto |{\bf x}|^2= R^2$. 
This is normally observed
by 
the condensate 
density 
going to zero 
linearly like $1-r^2\equiv1-R^2/R^2_{\rm b}$
near the border $R_{\rm b}$
(in the Thomas-Fermi approximation)  \cite{FD}. 
For $B$ near $B_c$ (or $T$ near $T_c$), however, 
the anomalous power $\delta$ will
lead to
the steeper
approach to zero $(1-r^2)^{2\beta}$ where
$2\beta\equiv\nu(D-2+\eta)=1-3\epsilon/10+\dots\approx0.7,$
plotted in Fig.~1, as will be shown immediately.
In addition, the central region is depleted. 
\begin{figure}[h]
\vspace{1.cm}
\hspace{-.6cm}
\unitlength.5pt
\hspace{-15em}
\begin{picture}(94.49,127.195)
\input epsf.sty
\def\dst{\displaystyle}
\def\IncludeEpsImg#1#2#3#4{\renewcommand{\epsfsize}[2]{#3##1}{\epsfbox{#4}}}
\put(-20,0){\IncludeEpsImg{94.49mm}{64.13mm}{0.300}{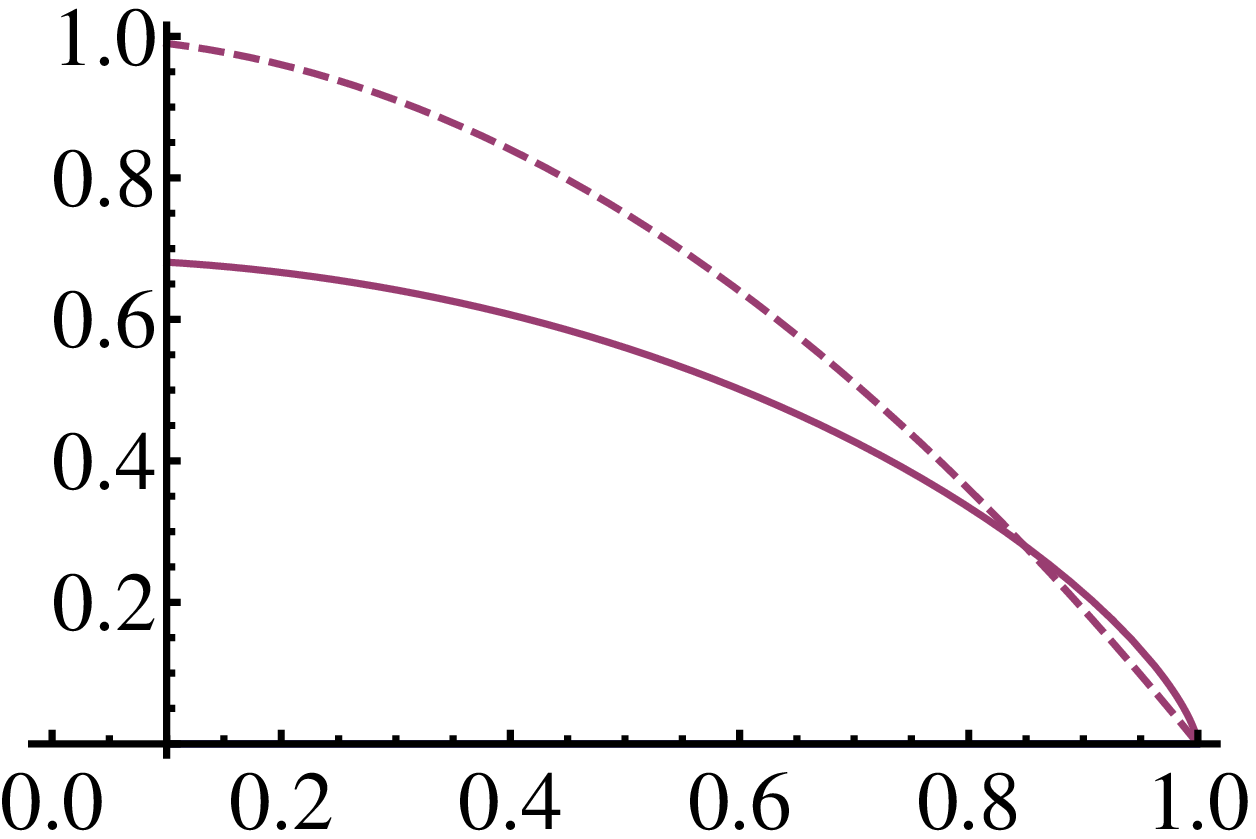}}
\put(220,0){\IncludeEpsImg{94.49mm}{64.13mm}{0.300}{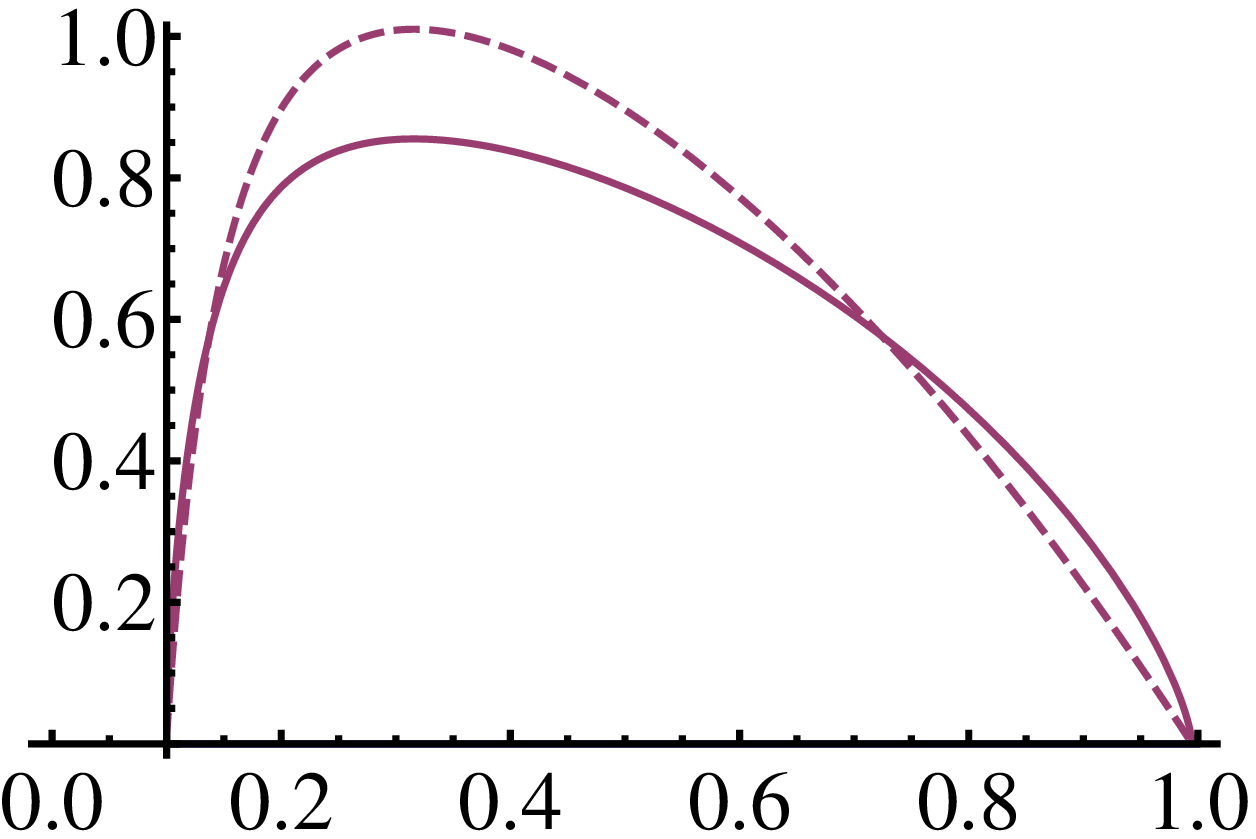}}
\put(78,-6){ $ r$}
\put(318,-6){ $ r$}
\put(50,111){\fsz GP}
\put(50,80){\fsz FGP}
\put(285,131){\fsz GP}
\put(285,105){\fsz FGP}
\put(48,40){\fsz $\rho=\Psi^*\Psi$}
\put(288,50){\fsz $\rho=\Psi^*\Psi$}
\end{picture}
\caption[]{Condensate density
from Gross-Pitaevskii equation
(\ref{@GP}) (GP,dashed) and its fractional version 
(\ref{@fGP} (FGP), both in Thomas-Fermi approximation
where the gradients are ignored. The FGP-curve shows a marked 
depletion of the condensate.
On the right hand, a vortex is included.
The zeros at $r\approx 1$will be smoothened by
the gradient terms in  (\ref{@GPP}).
}
\label{@f0}\end{figure}

Let us study the appearance of a reduced mass 
 $\hat m^2\propto (1-r^2)$
in the trap. The effective action will
introduce it into
 (\ref{@GPP}) in the form $\mu^{2-\eta}(\hat m^2)^{\nu(2-\eta)}f(|\Psi|^2/
(\hat m^2)^{2 \beta})$ with a Taylor series of $f(x)$
(note that $
\nu (2-\eta)= 1+\frac{\epsilon}5+\dots\approx1.3$).
For small $\hat m^2
$,
this may be resummed to a Widom type expression
 $[(\delta+1)/4\mu^\eta]g_c|\Psi|^{\delta-1}w(\hat m^2/|\Psi|^{1/\beta})$
\cite{KS22}. This explains the earlier-stated steeper falloff $|\Psi|^2\propto 
(\hat m ^2)^{2\beta}$ of the 
density profiles in Fig. 1.
The Widom function $w(\hat m^2/|\Psi|^{1/\beta})$ can be expanded as 1
plus a
power series in $(\hat m^2)^{\omega/2\nu}\propto
\xi^{-\omega}$ which contains
the Wegner 
critical exponent
$\omega\approx 0.8$ that governs the
{\it approach to scaling\/}  \cite{REMKIN}.
Thereby the
interaction term $|\Psi|^{\delta-1}$ is modified to $|\Psi|^{\delta-1}(1+{\rm const}\times 
\xi^{-\omega}|\Psi|^{-\omega \nu/\beta})$. Similarly, the kinetic terms $ (\hat{\bf p}^2)^{1-\eta/2}$ in (\ref{@GPP})
and (\ref{@fGP})
 approach the scaling limit like
$ (\hat{\bf p}^2)^{1-\eta/2}[1+{\rm const}\times\xi^{-\omega
}(\hat{\bf p}^2)^ {-\omega/2}+\dots]$ \cite{REMDEL}. 

A further observable phenomenon is
that 
the resonance frequency
of a forced collective 
oscillation will depend  on 
the field strength
$B$ near the Feshbach resonance \cite{VDLB}.

Summarizing we have seen that a many-body system with strong couplings between the constituents
satisfies
a more general form of the \sch{} equation, in which the momentum and the energy
appear with a power different from $\alpha=2$ and $\gamma=0$, respectively.
The associated Green function
can be written as a path integral over fluctuating
 time and space orbits that are functions of some pseudotime $s$.
This is a Markovian object, but non-Markovian in the physical time
$t$
that is related to $s$ by a stochastic differential equation
of the Langevin type.
The particle distributions
can  also be obtained by solving a
 Langevin type of
equation
in which the noise has correlation functions
whose probability distribution
is  specified.



{
{~\\
Acknowledgment:
I am grateful to P. Jizba, N. Laskin, M. Lewenstein, A. Pelster, and M. Zwierlein
for useful comments.
}}

~\\
{\bf Appendix}:
The lowest-order critical exponents
can be extracted directly
from the one-loop-corrected inverse Green function $G^{-1}(E,{\bf p})$
in $D=2+\epsilon$ dimensions after a
minimal subtraction of the $1/\epsilon$ -pole at \cite{ABBR}:
\begin{eqnarray}
E\!-\!{\bf p}^2\!+
a
\left(\sfrac{1}3{\bf p}^2\!-\!E\right)^{D-1}
\comment{
\left[\frac{\Gamma(1-D)}{
(\sqrt{8}\pi)^D(3!)^{D/2+1}}\!-\!\frac{1}{288 \pi^2 \epsilon}\right]
}
.
\label{@}\end{eqnarray}
For ${\bf p}=0$, this has a power
$-(-E)^{1-a\epsilon}$, so that $\gamma=a\epsilon$.
For $E=0$, on the other hand,
we obtain $(-{\bf p}^2)^{1-a\epsilon/3}$, so that
$(1-\gamma)/z-1\approx \gamma/3$.


\begin{thebibliography}{9}
\bibitem{remn}
A travelling  pedestrian
salesman is a Gaussian
random walker, as a jetsetter he becomes a L\'evy random walker.
\bibitem{METZ}
The relevant functional matrix is
$\langle{\bf x} |
(-\nablabf^2)^{\lambda/2}|{\bf x} '\rangle=\Gamma[-\lambda/2]^{-1}
\int d\sigma\, \sigma^{-\lambda/2-1}
{(4\pi \sigma)^{-D/2}}e^{R^2
/4\sigma}
$=
$\!\!{}^D\! c_\lambda R^{-\lambda-D}$, where
${}^D\!c_\lambda=2^{\lambda}\Gamma((D+\lambda)/2)/\pi^{D/2}\Gamma(-\lambda/2),$
and $R\equiv|{\bf x}-{\bf x}'|$. If $\lambda$ is close to an
even integer, it needs a small positive shift $\lambda\rightarrow\lambda_+\equiv\lambda+\epsilon$ and we can replace $ \epsilon R^{\epsilon-1}/2 $ by 
$\delta(R)=S_DR^{D-1}\delta^{(D)}({\bf R})$.
For  $A>0$
 we have $|{\bf x}'
|^{-A}= {}^D\!c_{\lambda_A}^{-1}\langle{\bf x}' |
(-\nablabf^ 2)^{\lambda_A/2}|{\bf 0} \rangle$ with
$\lambda_A\equiv A-D$, 
so that we find
$\int d^Dx'\langle{\bf x} |(-\nablabf^ 2)^{\lambda/2}|{\bf x}' \rangle |{\bf x}'|^{-A}=
{}^D\!c_{\lambda_A}^{-1}\langle{\bf x}|(-\nablabf^ 2)^{(\lambda+A-D)/2}
|{\bf 0}\rangle =
{}^D\!c_{\lambda+A-D}{}^D\!c^{-1}_{\lambda_A}|{\bf x}
|^{-A-\lambda}$.
\bibitem{REM1}
This technique is explained in Chapters~12 and 19 of Ref.~\cite{PI}.
The pseudotime $s$ resembles the so-called
 Schwinger proper time used in relativistic physics.
\bibitem{REMAN}

There should be  no danger of confusing the fluctuating
noise variable $\eta$
in this equation  with the constant
critical exponent $\eta$ in (\ref{@PropaGAL2}).
\bibitem{REMAR}
The decimal numbers are from seven-loop calculation in $D=3$ dimensions 
in Table 20.2 of Ref.~\cite{KS22}.
\end{thebibliography}

\end{document}